\documentclass{eptcs}
\providecommand{\event}{P\&C 2018} 

\usepackage{breakurl} 
\usepackage{parskip,setspace}
\usepackage{amsmath,amssymb,amsthm}
\usepackage{listings}
\usepackage{textcomp}
\usepackage[pdftex]{graphicx}
\usepackage[utf8]{inputenc}
\usepackage[T1]{fontenc}

\newtheorem{Theorem}{Theorem}

\theoremstyle{definition}

\newcommand{\vect}[1]{\mathbf{#1}}
\newcommand{\xvec}{\vect{x}}

\DeclareMathOperator*{\argmin}{\arg\min}

\usepackage{hyperref}

\title{A Hybrid Quantum-Classical Paradigm to Mitigate Embedding Costs in Quantum Annealing---Abridged Version\thanks{A significantly expanded version is available at \href{https://arxiv.org/abs/1803.04340}{arXiv:1803.04340 [cs.DS]}.}}
\author{Alastair A.\ Abbott
\institute{Univ.\ Grenoble Alpes, CNRS,
Grenoble INP, Institut N\'eel\\ 38000 Grenoble, France}
\and
Cristian S.\ Calude \qquad Michael J.\ Dinneen \qquad Richard Hua
\institute{Department of Computer Science,
 University of Auckland\\ Private Bag 92019, Auckland, New Zealand}
}

\def\titlerunning{A Hybrid Quantum-Classical Paradigm to Mitigate Embedding Costs in Quantum Annealing}
\def\authorrunning{A.\ A.\ Abbott, C.\ S.\ Calude, M.\ J.\ Dinneen \& R.\ Hua}

\begin{document}

\maketitle

\begin{abstract}
	Quantum annealing has shown significant potential as an approach to near-term quantum computing.
	Despite promising progress towards obtaining a quantum speedup, quantum annealers are limited by the need to embed problem instances within the (often highly restricted) connectivity graph of the annealer.
	This embedding can be costly to perform and may destroy any computational speedup. 
	Here we present a hybrid quantum-classical paradigm to help mitigate this limitation, and show how a raw speedup that is negated by the embedding time can nonetheless be exploited in certain circumstances.
	We illustrate this approach with initial results on a proof-of-concept implementation of an algorithm for the dynamically weighted maximum independent set problem.	
\end{abstract}
	
\section{Introduction}

Quantum computation has the potential to revolutionise computer science, and as a consequence has received a great deal of attention from theorists and experimentalists alike.
Although much progress has been made through the concerted efforts of the community, we are still some distance from being able to build sufficiently large-scale universal quantum computers to realise this potential~\cite{Ladd:2010aa}.

More recently, however, significant progress has been made in the development of special-purpose quantum computers.  
This has been driven by the realisation that, by dropping the requirement of being able to efficiently simulate arbitrary computations and relaxing some of the constraints that make large-scale universal quantum computing difficult, such devices can be more easily engineered and scaled.  
With this approach it may be possible to exploit some of the capabilities of quantum computation to obtain lesser, but nevertheless practical, advantages in real-world applications.  
Quantum annealers, which solve particular optimisation problems, exemplify this approach, and significant progress has been made in recent years towards engineering moderately large-scale such devices~\cite{Johnson:2011aa}.  
This approach has been pursued particularly zealously by D-Wave, who have developed quantum annealers with upwards of 2000 qubits~\cite{dwavesys2017} and  are thus of sufficient size to tackle problems for which their performance can meaningfully be compared to classical computational approaches.

In this paradigm, however, it is a subtle problem to compare the performance of quantum solutions to classical ones, since the focus is on obtaining real-world gains in domains where heuristics tend to be at the core of the best classical approaches.
Indeed, this issue is at the heart of recent debate surrounding the performance of D-Wave machines~\cite{Cho:2014aa,Shin:2014aa}.
In particular, instead of focusing on asymptotic analyses, one must compare the performance of classical and quantum devices
empirically.
But performing benchmarking fairly is difficult, especially when there is often debate as to which classical algorithm should be taken for comparison~\cite{King:2015ab,King:2015aa,Ronnow:2014aa}.

In this paper, motivated by the need to take into account the cost of  classical processing in benchmarking quantum annealers, we propose a hybrid quantum-classical approach for developing algorithms that mitigates the cost of this processing.
We focus on D-Wave's quantum annealers where the process involves a costly classical ``embedding'' stage  which maps an arbitrary problem instance into one compatible with D-Wave's limited connectivity constraints.
We then formulate a generic hybrid approach that mitigates this cost allowing any advantage present to be accessed more directly~\cite{Calude:2015aa}.
The embedding problem  is time-consuming,  and experimental studies indicate that its quality can have strong effects on 
performance~\cite{jobshopschedulingproblem2016,maxindepsetqubo2017}.

To illustrate this generic framework for hybrid computing we propose, we  present a hybrid algorithm based around a D-Wave solution to the maximum-weight independent set (MWIS) problem.
We present an overview of the results of an initial proof-of-principle implementation of this algorithm, showing a large improvement of the hybrid algorithm over a more standard quantum annealing approach, as well as comparing it to a classical algorithm.

\section{D-Wave's quantum annealing framework}

\subsection{Quantum annealing and quadratic unconstrained Boolean optimisation}

Quantum annealing is a finite temperature implementation of adiabatic quantum computing~\cite{Farhi:2000aa}, in which the optimisation problem to be solved is encoded into a Hamiltonian $H_p$ (the quantum operator corresponding to the system's energy) such that the ground state of $H_p$ corresponds precisely to the solution to the problem (or one of them, if many exist).
The computer is initially prepared in the ground state of a Hamiltonian $H_i$, which is then slowly evolved into the target Hamiltonian $H_p$.
This computation can be described by the time-dependent Hamiltonian $H(t)=A(t)H_i + B(t)H_p$ for $0\le t \le T$, where $A(0)=B(T)=1$ and $A(T)=B(0)=0$.
$T$ is called the annealing time and, for D-Wave machines, the functions $A$ and $B$ give a close to a linear transition from $H_i$ to $H_p$~\cite{Johnson:2011aa}.
If the computation is performed sufficiently slowly, the Adiabatic Theorem guarantees that the system will remain in a ground state of $H_p$ throughout the computation and the final state will thus correspond to an optimal solution to the problem at hand~\cite{Farhi:2000aa}.

Quantum annealers implement specific, simple classes of Hamiltonians, such as the two-dimensional Ising spin Hamiltonians realised by D-Wave devices.
This restriction means that D-Wave annealers are capable of solving natively the \emph{Quadratic Unconstrained Boolean Optimisation (QUBO) problem}~\cite{Choi:2008aa}. 
The QUBO problem is the task of finding the input $\vect{x}^*$ that minimises a quadratic objective function of the form $f(\vect{x}) = \vect{x}^{T}Q\vect{x}$, where $\vect{x}=(x_1,\dots,x_n)$ is a vector of $n$ binary variables and $Q$ is an upper-triangular $n \times n$ matrix of real numbers:
\[\vect{x}^* = \argmin_{\xvec}\vect{x}^{T}Q\vect{x}  = \argmin_{\xvec} \sum_{i\leq j} x_i Q_{(i,j)} x_j, \text{ where } x_i \in \{0,1\}.\]
In the quantum annealing model of the QUBO problem, each $x_i$ corresponds to a qubit while $Q$ defines the problem Hamiltonian $H_p$.
Crucially, the nonzero terms $Q_{(i,j)}$ (for $i\neq j$) correspond to couplings between qubits and induce a graph $G_L=(V_L,E_L)$ called the \emph{logical graph} representing interactions between qubits; the qubits $V_L$ in which the QUBO problem is represented over are called the \emph{logical qubits}.

\subsection{Hardware constraints and embeddings}\label{sec:Chimera}

The comparative ease in engineering devices which naturally solve the QUBO problem has been crucial for the recent experimental success of quantum annealing.
Still,  it remains exceedingly difficult to control interactions between qubits that are not physically near to one another, and as a result it is not possible to implement directly any instance of the QUBO problem.
Instead, the couplings possible on a quantum annealer are specified by a \emph{physical graph} $G_P=(V_P,E_P)$, where $V_P$ is the set of \emph{physical qubits} on the device, and an edge $\{i,j\}\in E_P$ signifies that qubits $i$ and $j$ can be physically coupled~\cite{Choi:2008aa}.

The physical graphs implemented on D-Wave's devices are \emph{Chimera graphs} $\chi_{k}$, which are $k\times k$ grids of $K_{4,4}$
graphs~\cite{dwavebroadcast2016}. 
Specifically, each qubit is coupled with four other qubits in the same $K_{4,4}$ block and two qubits in adjacent blocks (except for qubits in blocks on the edge of the grid, which are coupled to a single other block).
The Chimera graph is crucially relatively sparse and near-to-planar, with qubits separated by paths of length up to $2k$.
Since the logical graph $G_L$ for a QUBO problem instance $Q$ will not, in general, be a subgraph of the physical graph $G_P=\chi_k$, the problem instance on $G_L$ must be mapped to an equivalent one on $G_P$.
This process involves two steps: first, $G_L$ must be \emph{embedded} in $G_P$, and secondly the weights of the QUBO problem (i.e., the non-zero entries in $Q$) must be adjusted so that valid solutions on $G_P$ are mapped to valid solutions on $G_L$.

The embedding stage amounts to finding a \emph{(minor) embedding} of $G_L$ into $G_P$~\cite{Choi:2008aa}, i.e., an embedding function $f:V_L\to 2^{V_P}$ such that
i)  the sets of vertices $\{ f(v) \mid v \in V_L\}$ are disjoint, ii)
 for all $v \in V_{L}$, there is a subset of edges $E' \subseteq E_{P}$ such that $G' = (f(v), E')$ is connected and iii)
 if $\{u,v\} \in E_{L}$, then there exist $u',v' \in V_{P}$ such that $u' \in f(u)$, $v' \in f(v)$ and $\{u',v'\}$ is an
edge in $E_{P}$.
The problem of finding a minor embedding is itself computationally difficult~\cite{Choi:2008aa}.
The embedding process may thus, in light of its computational difficulty, contribute significantly to the time required to solve a problem in practice. 
Currently, the standard approach to finding such an embedding is to use heuristic algorithms.

\subsection{Benchmarking quantum annealers}\label{sec:measure_time}

It is not generally believed that an exponential speedup is possible for NP-hard problems such as the QUBO problem~\cite{Aaronson:2010aa}, and there has been much debate over whether or not quantum annealing provides any such speedup in practice~\cite{Calude:2015aa,Ronnow:2014aa}.
Indeed, there is disagreement over what exactly constitutes a quantum ``speedup'' and how to determine if there is one. 
In this paper we will focus primarily on the empirical run-time performance in investigating whether a quantum speedup is present, rather than the (empirically estimated) scaling performance of quantum algorithms.

Good benchmarking needs to make use of fair and comprehensive metrics to determine the running time of both classical and quantum algorithms for a problem.
In particular these need to properly take into account not only the ``wall-time'' of different stages of the quantum algorithm, but also its probabilistic nature.
To understand how this can be done, we first need to outline the different stages of the quantum annealing process~\cite{King:2014aa}:
1)  \emph{Programming:} The problem instance is loaded onto the annealing chip (QPU), which takes time $t_{\text{prog}}$;
2)  \emph{Annealing:} The quantum annealing process is performed and then the physical qubits are measured to obtain a solution; this takes time
$t_a$;
3)  \emph{Repetition:} Step 2 is performed $k$ times to obtain $k$ potential solutions.
The \emph{quantum processing time} (QPT) is thus $t_{\text{proc}}=t_{\text{prog}}+k\,t_a.$
With these considerations on hand, a relatively fair and robust way to measure the quantum processing time is  the ``time to solution'' (TTS) metric~\cite{Boixo:2014aa,Ronnow:2014aa}, which is based on the expected number of repetitions needed to obtain a valid solution with probability $p$ (one typically takes $p=0.99$).
If the probability per annealing sample of obtaining a solution is $s$ (which can be estimated empirically), then this is calculated as 
$k_{99}=\frac{\log(1-p)}{\log(1-s)}\raisebox{0.7mm}{,}$ and the quantum processing time is thus calculated with this $k$ as $t_{\text{proc}}=t_{\text{prog}}+k_{99}\,t_a$.
Existing investigations have primarily focused on comparing directly the QPT with the processing time of a classical algorithm in order to look for what we call a ``raw quantum speedup''.
However, it is essential to realise that the time used by the QPU and measured by the QPT refers only to a subset of the processing required to solve a given problem instance using a quantum annealer.
Specifically, a complete quantum algorithm for a problem instance $P$ involves, as a minimum requirement, the following steps:
\begin{enumerate}
\item \emph{Conversion:} The problem instance $P$ must be converted into a QUBO instance $Q(P)$, typically via a polynomial-time reduction taking time $t_{\text{conv}}$.
\item \emph{Embedding:} The QUBO problem $Q(P)$ must be embedded into the Chimera hardware graph taking time $t_{\text{embed}}$.
\item \emph{Pre-processing:} The embedded problem is pre-processed, which involves calculating (appropriately scaled) weights for the embedded QUBO problem, taking time $t_{\text{pre}}$.
\item \emph{Quantum processing:} The annealing process is performed on the QPU, taking time $t_{\text{proc}}$.
\item \emph{Post-processing:} The samples are post-processed to choose the best candidate solution, check its validity, and perform any other post-processing methods to improve the solution quality~\cite{King:2014aa,Pudenz:2016aa}
 taking time $t_\text{post}$. The QUBO solution must finally be converted back to a solution for the original problem $P$.
\end{enumerate}
The total processing time is thus
\begin{equation}\label{eqn:Tq}
T_Q=t_{\text{conv}}+t_{\text{embed}}+t_\text{pre}+t_{\text{proc}}+t_\text{post}.
\end{equation}
The realisation that these other steps must be included in the analysis is emphasised by the fact that in   practical  problems the embedding time often dominates the time used by the annealer itself.
Previous investigations have largely avoided this by focusing on artificial problems ``planted'' in the Chimera graph so that no embedding is necessary~\cite{Boixo:2014aa,Denchev:2016aa,Hen:2015aa,Ronnow:2014aa,Shin:2014aa}. 
Although finding a raw speedup in such situations is clearly a necessary condition for a quantum speedup, it is not sufficient for it to be present in   practical  problems.

To properly benchmark quantum annealing it is necessary to also compare fairly the quantum annealer to a suitable classical algorithm.
Indeed, much of the controversy regarding potential speedups with quantum annealing has been due to the fact that quantum annealers have been compared against simulated annealing or simulated quantum annealing.
Although such studies certainly have merit and such a speedup is certainly a necessary condition for a real quantum speedup, it has repeatedly been pointed out that classical annealing techniques are generally far from optimal and any observed speedups have disappeared when better classical algorithms were used~\cite{Denchev:2016aa}.
In~\cite{Ronnow:2014aa}, this type of quantum speedup has been termed a ``limited speedup''.
Ideally, one should instead compare annealing against the \emph{best available} classical algorithm for the problem to find a ``potential quantum speedup''.

\section{Hybrid quantum-classical computing}

Much of the previous effort towards determining whether or not quantum annealing can, in practice, provide a computational speedup has focused on determining the existence of a \emph{raw quantum speedup}, which does not take into account the associated classical  processing that is inseparable from a quantum annealer.
Such a raw speedup is certainly a necessary condition for practical quantum computational gains, and its study is therefore well justified. However, even if there is a raw speedup there are many reasons why this might not translate into a \emph{practical} quantum speedup.

A practical speedup is possible for a problem if we are able to give a quantum algorithm such that $T_Q<T_C$, where $T_C$ is the classical processing time  for the best available classical algorithm for the problem.
From the definition of $T_Q$ in \eqref{eqn:Tq}, it is clear than, even if $t_{\text{proc}}<T_C$, the conversion, embedding and pre/post-processing may provide obstructions to obtaining a practical speedup.
In practical terms, the pre- and post-processing tend to add relatively minor (or controllable) overheads, but the conversion and embedding costs pose more fundamental problems.

These difficulties in turning a raw quantum speedup into a practical advantage for   practical  problems have led to significant interest in ``hybrid classical-quantum'' approaches (also called ``quassical'' computations by Allen~\cite{Calude:2015aa}).
Hopefully,  combining quantum annealing with classical algorithms may allow otherwise inaccessible speedups to be exploited.
Several such hybrid approaches have aimed to overcome the resource limitation arising from the fact that   practical  problems typically require more qubits than are available on existing devices (as a result of the expansion in number of variables during the conversion stage discussed above)~\cite{McClean:2016aa,Tran:2016aa}.

\subsection{Hybrid computing to negate embedding costs}
\label{sec:hybridAlgGen}

Although hybrid approaches have also looked at improving the robustness and quality of embeddings~\cite{Vinci:2015aa}, to the best of our knowledge such approaches have not been used to try and mitigate the cost of performing the embedding itself, which, we recall, is often prohibitive to any speedup.
In this paper we propose a general hybrid approach to tackle precisely this problem.
In particular we aim to show how a raw speedup that is negated by the embedding time (i.e., in particular when $t_{\text{proc}}<T_C$ but $T_Q>T_C$) can nonetheless be exploited to give a practical speedup to certain computational problems.

Our approach is motivated by another hybrid quantum-classical algorithmic proposal which predates the rise of quantum annealing and was introduced with the aim of exploiting Grover's algorithm---the well-known black-box algorithm for quantum unordered database search~\cite{Grover:1996aa}---in practical applications~\cite{Lanzagorta:2005aa}.
The crucial condition for such a problem to be amenable to this hybrid approach is that \textit{the repeated calls to the quantum annealer should be made with the same logical graph embedding}, or at least \textit{permit an efficient method to construct the embedding for one call from the previous ones.}
If this condition is satisfied, the cost of the embedding, $t_{\text{embed}}$, can thus be spread out over the several calls, allowing a raw quantum speedup to be exploited.

In order to see how this hybrid approach can help exploit a quantum speedup, we will consider the following general description of a quantum annealing algorithm based on the hybrid approach described above (a more precise analysis would necessarily depend in part on the algorithm in question): 
some initial classical processing is performed, the embedding of a logical graph into the physical graph is computed, $m$ instances of a QUBO problem are solved on a quantum annealer, with some classical pre- and post-processing occurring between instances, and some final classical computation is optionally performed.
More formally, let us call the overall problem the hybrid algorithm solves $R$, and the $m$ problem instances that must be solved to do so, $P_1,\dots,P_m$.
Recall that the time to solve a single instance $P_i$ on an annealer is $T_Q(P_i)$. As we noted earlier this is, in practical situations, generally dominated by the cost of the embedding and the quantum processing, so $T_Q(P_i)$ can be approximated, for simplicity, as
\begin{align}
T_Q(P_i) &= t_{\text{conv}}(P_i) + t_{\text{embed}}(P_i) + t_\text{pre}(P_i) + t_{\text{proc}}(P_i) + t_\text{post}(P_i)
\approx t_{\text{embed}}(P_i) + t_{\text{proc}}(P_i),
\end{align}
where we have explicitly included the dependence on the problem instance.
The hybrid algorithm will thus take time
\[
T_H(R) 
\approx t_1(R) + t_{\text{embed}}(P_1) + \sum_i\big(t_{\text{proc}}(P_i) + t_2(P_i)\big)\notag\\
\approx t_1(R) + t_{\text{embed}}(P_1) + \sum_i t_{\text{proc}}(P_i),
\]
where $t_1(R)$ encapsulates any initial and final classical processing associated with combining the solutions $P_i$, and $t_2(P_i)$ is the time of the  classical calculation associated with each iteration, which we have assumed to be small compared to $t_{\text{proc}}(P_i)$ since this should simply encompass minor pre- and post-processing between annealing runs, and thus be negligible if the problem is amenable to the hybrid approach.
Note that we have made use of the assumption that $t_\text{embed}(P_1)\approx t_\text{embed}(P_i)$ for $i>1$, which is a criterion in the suitability of a problem for this hybrid approach.

We note immediately that a standard approach with a quantum annealer, performing the embedding for each instance $P_i$, would take time
$T_\text{std}(R) = t_1(R) + \sum_i\big(t_{\text{embed}}(P_i) + t_{\text{proc}}(P_i)\big).$
Thus, since in practice $t_{\text{embed}}$ is comparable, if not larger, than $t_{\text{proc}}$, we already have $T_H(R) \ll T_\text{std}(R).$
Although this conclusion may seem somewhat trivial, it is important in that it shows already how annealing can provide much larger practical gains for such complex algorithmic problems.
More importantly,  \textit{it may allow a raw quantum speedup to be exploited practically.}

It is, of course, possible that for certain problems a much more efficient classical algorithm exists for solving $R$ when $m$ is large enough (e.g., there might be an efficient way to map solutions of $P_i$ to $P_j$). Such problems are thus not suitable for such a hybrid approach, and so are not of particular interest to us.
Nonetheless, generally a classical algorithm for $R$ may be more intelligent than the standard approach as certain, presumably minor,
parts of the computation are likely to be common to solving several $P_i$.
Specifically, we can thus rewrite $T_C(P_i)=t_3(P_i) + t_4(P_i)$, where $t_3$ is small compared to $t_4$.
The best classical algorithm can then, rather generally, be considered to take time 
\[T_C^{\text{best}}(R) =t_5(R) + t_3(P_1) + \sum_i t_4(P_i)
= t_6(R) + \sum_i t_4(P_i),\]
\noindent where $t_6(R)=t_5(R) + t_3(P_1)$.
Crucially, unless the raw quantum speedup is small, we will also have $t_{\text{proc}}(P_i)<t_4(P_i)$.
It is thus easy to see that, 
\emph{for large enough $m$ (i.e., number of $P_i$ to be solved), we will have $T_H(R) < T_C^{\text{best}}(R)$},
and thus the raw quantum speedup will translate into an absolute speedup for the hybrid algorithm.

\section{Case study: Dynamically weighted maximum-weight independent set}

To illustrate the proposed hybrid approach, we discuss in detail a concrete example both from a theoretical and experimental viewpoint.

\subsection{(Dynamically weighted) Maximum-weight independent set} \label{sec:MWIS}

The core of the problem is the maximum-weighted independent (MWIS) set problem.
Recall that an \emph{independent set} $V'$ of vertices of a graph $G=(V,E)$ is a set $V' \subseteq V$ such
that for all $\{u,v\} \in E$ we have $\{u,v\} \not\subseteq V'$.  

\begin{samepage}
\noindent{\bf Maximum-Weight Independent Set (MWIS) Problem}:\\[1.5ex]
\begin{tabular}{ll}
\emph{Input:} & A graph $G=(V,E)$ with positive vertex weights $w:V \rightarrow \mathbb{R}^+$. \\
\emph{Task:} & Find  an independent set $V' \subseteq V$ 
such that maximises
$\sum_{v \in V'} w(v)$ \\
& over all independent sets of $G$.
\end{tabular} 
\end{samepage}

The general MWIS problem is NP-hard since it encompasses, by restriction, the well-studied non-weighted version~\cite{GJ79}. 
One should note, however, that for graphs of bounded tree-width, the MWIS problem is polynomial-time solvable using standard 
dynamic programming techniques (see~\cite{Marx10}).

Although the MWIS can be readily transformed into a QUBO problem (as we show below), by itself it is not directly suitable for the hybrid approach we proposed. 
However, a simple variation that we propose here is indeed suitable.

\begin{samepage}
\noindent{\bf Dynamically Weighted Maximum-Weight Independent Set (DWMWIS) Problem}:\\[1.5ex]
\begin{tabular}{ll}
\emph{Input:} & A graph $G=(V,E)$ with a set of weight functions $W=\{w_1, w_2, \ldots, w_m\}$ \\ 
& where $w_i:V \rightarrow
\mathbb{R}^+$ for $1 \leq i \leq m$. \\
\emph{Task:} & Find independent sets $V_i \subseteq V$ that maximise 
$\sum_{v \in V_i} w_i(v)$ for each $1 \leq i \leq m$.
\end{tabular} 
\end{samepage}

This problem is to solve the MWIS problem on $G$ for each of the $m$ weight assignments $w_i\in W$.
For $m=1$ 
we obtain again the MWIS problem, but for larger $m$ 
the problem is suitable for our hybrid approach.

\subsection{Quantum solution}\label{sec:quantumSoln}

We now provide a QUBO formulation for the MWIS Problem.
Fix an input graph $G=(V,E)$ with positive vertex weights $w:V \rightarrow \mathbb{R}^+$.
Let $W=\max \{w(v) \mid v \in V \}$ and let $S>W$ be a ``penalty weight''.
We build a QUBO matrix of dimension $n=|V|$ such that:

\begin{equation}\label{MWISQUBO}
Q_{(i,j)} = \left\{
\begin{tabular}{cl}
0,       & \mbox{if } $i>j \mbox{ or } \{i,j\} \not\in E,$ \\
$-w(v_i)$, & \mbox{if } $i=j,$ \\
$S$,       & \mbox{if } $i<j \mbox{ and } \{i,j\} \in E.$ \\
\end{tabular}    
\right.
\end{equation}

\medskip

\begin{Theorem}[\cite{Abbott18}]
The QUBO formulation given in~\eqref{MWISQUBO} solves the MWIS Problem.
\end{Theorem}

In order to adapt the MWIS solution above to the DWMWIS problem, note that the non-zero entries of the QUBO formulation~\eqref{MWISQUBO} depend only on the structure of the graph and not on the weight function $w$.  
Thus, in order to solve the DWMWIS problem, for each weight assignment $w_i$ the same embedding of the graph into the D-Wave physical graph can be used, meaning that a hybrid algorithm based around the MWIS solution above can readily be implemented.
More specifically, following the hybrid algorithm described in Section~\ref{sec:hybridAlgGen} for instances $P_1, \ldots, P_m$ (where each $P_i$ uses weight function $w_i$), we perform the embedding once (entailing a time $t_{\text{embed}}(P_1)$) and then solve the MWIS problem for each weight assignment $w_i$ (taking times $t_{\text{proc}}(P_i)$) using the QUBO solution outlined above.

\subsection{Classical baseline}

The main objective of studying the DWMWIS example in detail is to exhibit experimentally the advantage that the hybrid approach can provide over a standard annealing-based approach.
Nonetheless, it is helpful to further compare this to the performance of a classical baseline algorithm for comparison and to help highlight this advantage, even if we do not necessarily expect to see an absolute quantum speedup from the hybrid algorithm.

To this end, for a given input graph $G=(V,E)$ with positive vertex weights $w: V \rightarrow \mathbb{R}^{+}$, we construct a Binary Integer Programming (BIP)  instance with $n=|V|$ binary variables as follows.
To each vertex $v_i$ in $G$ we associate the binary variable $x_i$, and for notational simplicity we will denote the collection of variables $x_i$ by a binary vector $\xvec = (x_0, x_1, \cdots, x_{n-1})$.
We thus have the BIP problem instance:
\begin{equation} \label{IP}
\begin{array}{ll@{}l}
\text{maximise }  \displaystyle\sum\limits_{v_i \in V} w(v_i)x_{i} \\
\text{subject to } \displaystyle x_i + x_j \leq 1 \mbox{ for all }  \{v_i,v_j\} \in E.
\end{array}
\end{equation}

Each constraint in~\eqref{IP} enforces the property that no adjacent vertices are chosen in the independent set while the objective function ensures an independent set with maximum sum value is chosen. 
Assuming we have the binary vector $\xvec$ which yields the optimal value of objective function~\eqref{IP}, we take $D(\xvec) = \{v_i \mid x_i=1\}$ to be the set of vertices selected as the maximum weighted independent set.
\\
\begin{Theorem}[\cite{Abbott18}]
The BIP formulation given in (\ref{IP}) solves the MWIS problem.
\end{Theorem}

The classical baseline used in the analysis is based on an implementation of the BIP formulation in Sage
Math~\cite{sagemath}, which has a well developed and optimised Mixed Integer Programming library.
To ensure that a fair comparison with the hybrid algorithm is possible, we formulate the classical algorithm for the overall DWMWIS problem such that \emph{the set of constraints in the BIP formulation is only computed once}. 

\subsection{Experimental definition and procedure}

To study the performance of the hybrid DWMIWS algorithm in a practical setting, we made use of a D-Wave 2X quantum annealer with $1098$ active physical qubits~\cite{DWave2X} to compare the performance of three algorithms on a selection DWMWIS problem instances: the ``standard'' quantum algorithm, in which the embedding is re-performed for each weight assignment; the hybrid DWMWIS algorithm; and the classical BIP-based solution described above.
We present here a summary of the experimental procedure and results; a more detailed presentation and analysis is available in an extended version of the paper~\cite{Abbott18}.

To this end we analyse the algorithms on a range of different graphs, in particular choosing $155$ graphs from a variety of common graph families with between 2 and 126 vertices.
Each graph was used to generate a single DWMWIS problem instance with $m=100$ weight assignments, each randomly generated as floating point numbers rounded to 2 decimal places within the range $[0.0,1.0)$. 
The problem instances were generated as standard adjacency list representations using SageMath~\cite{sagemath} with random weights.
The same procedure is used for the ``standard'' quantum algorithm, except the cost of the embedding is incurred for each weight assignment.

Since we are primarily interested in negating the impact of the embedding process in general applications, we made use of D-Wave's heuristic embedding algorithm~\cite{DWave:2013aa} to embed each logical graph in the physical graph.
Each graph was embedded 10 times to estimate $t_\text{embed}$ for each problem instance.
Finally, our tests were run with D-Wave's post-processing optimisation enabled.
While this adds a small overhead in time, this is well within the spirit of hybrid quantum-classical computing, and allowed us to solve more problems.
This post-processing method processes small batches of samples while the next batch is being processed~\cite{DWave:postprocessing}.
This ensures that it only contributes a constant overhead in time for each MWIS problem instance \emph{independent of the number of samples (and thus $k_{99}$)}.

\subsection{Results and analysis}

\begin{figure}[h]
\begin{center}
\begin{tabular}{cc}
\includegraphics[width=0.5\textwidth]{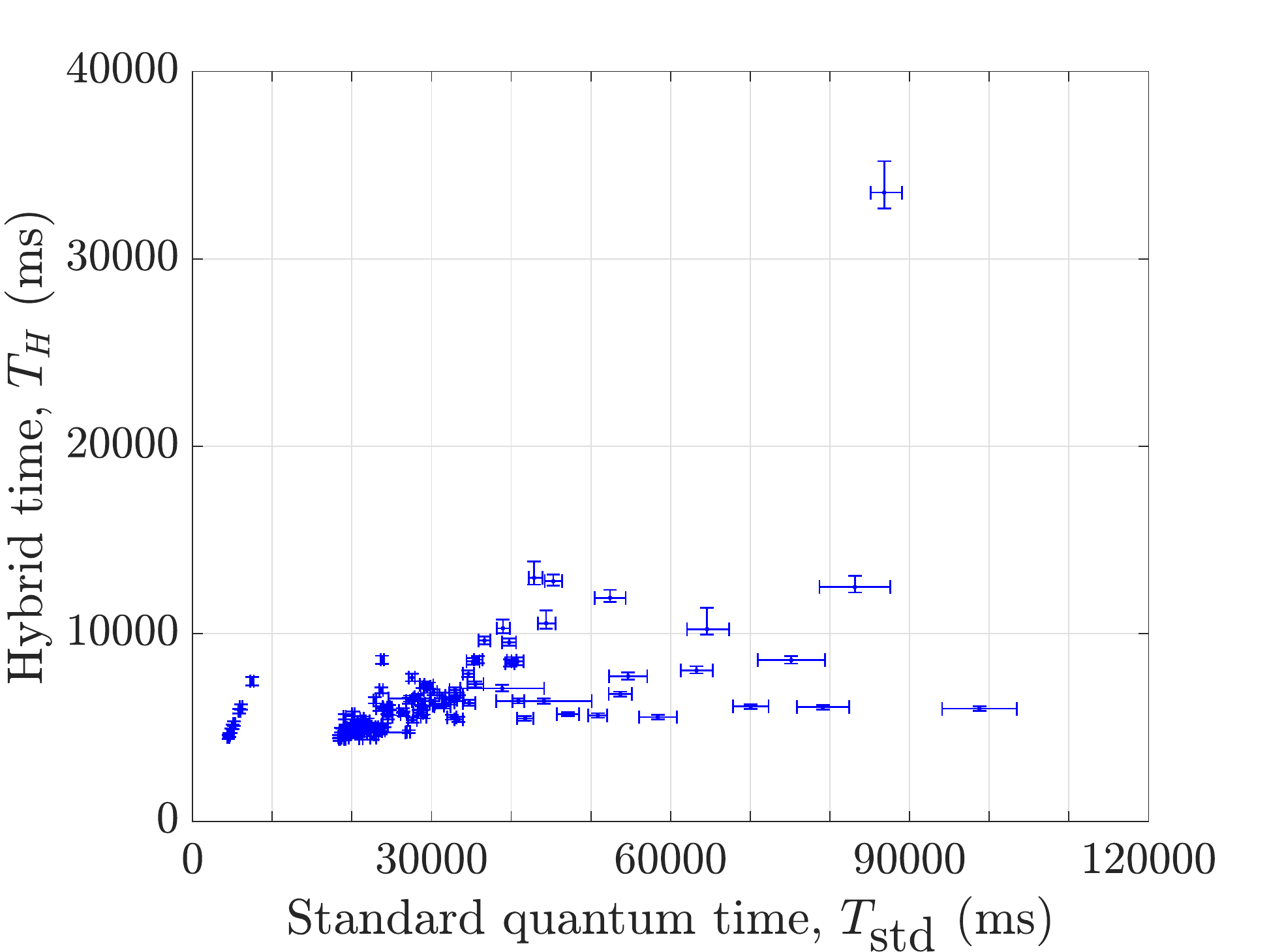} & \hspace{-5mm} \includegraphics[width=0.5\textwidth]{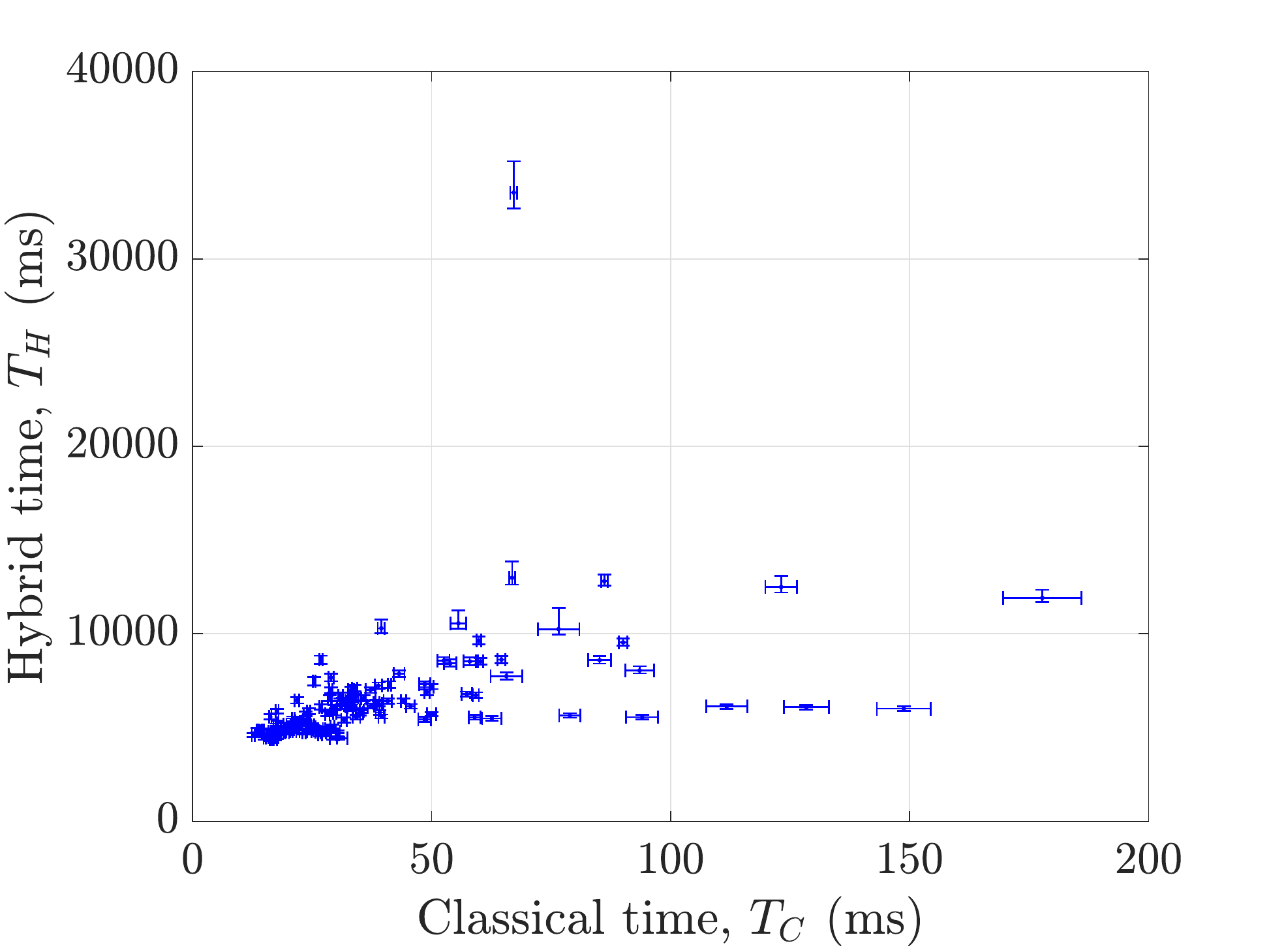}\\
(a) & (b)
\end{tabular}
\caption{Plots of (a) an upper bound for $T_\text{std}$ against $T_H$; and (b) $T_C$ against $T_H$ for each DWMWIS problem instance. All times are in ms.}\label{fig:TH_v_Tstd_v_TC}
\end{center}
\end{figure}

For each DWMWIS problem instance (i.e., for each graph $G$) the times $T_H$ and $T_\text{std}$ were calculated, following the approach described in Section~\ref{sec:hybridAlgGen}, as 
\begin{align*}
		T_H &=t_\text{embed} + \sum_i \big(t_\text{prog}(P_i) + k_{99}(P_i)t_\text{anneal} + t_\text{post}(P_i)\big),\\
	T_\text{std} &= \sum_i \big(t_\text{embed} + t_\text{prog}(P_i) + k_{99}(P_i)t_\text{anneal} + t_\text{post}(P_i)\big),
\end{align*}
where $k_{99}(P_i)$ is the $k_{99}$ value for weight assignment $w_i$ and $t_\text{anneal}=309\mu$s.
Both $t_\text{prog}(P_i)$ and $t_\text{post}(P_i)$ are of the order of $20$ms.
Note that the processing time $t_\text{proc}$ defined earlier is, for this approach to the DWMWIS problem, given by $t_\text{proc}=t_\text{prog}(P_i) + k_{99}(P_i)t_\text{anneal} + t_\text{post}(P_i).$
The classical time $T_C$ was taken as the processor time for the classical algorithm described above.
The results are summarised in Figures \ref{fig:TH_v_Tstd_v_TC}(a) and \ref{fig:TH_v_Tstd_v_TC}(b), which show how the hybrid times $T_H$ compare to both $T_\text{std}$ and $T_C$. 
Error bars are calculated from the observed variation in $t_\text{embed}$, the number of optimal solutions found $N_\text{opt}$, and the post-processing time $t_\text{post}$.
Of these, the error in $t_\text{post}$ is the dominant factor, and largely arises from the uncontrollability of the post-processing environment, which is performed remotely within the D-Wave processing pipeline.
However, this variation did not result in any significant variation in success probability of the annealing, so it seems the amount of post-processing performed was constant.

\begin{figure}[h]
\begin{center}
\includegraphics[width=0.5\textwidth]{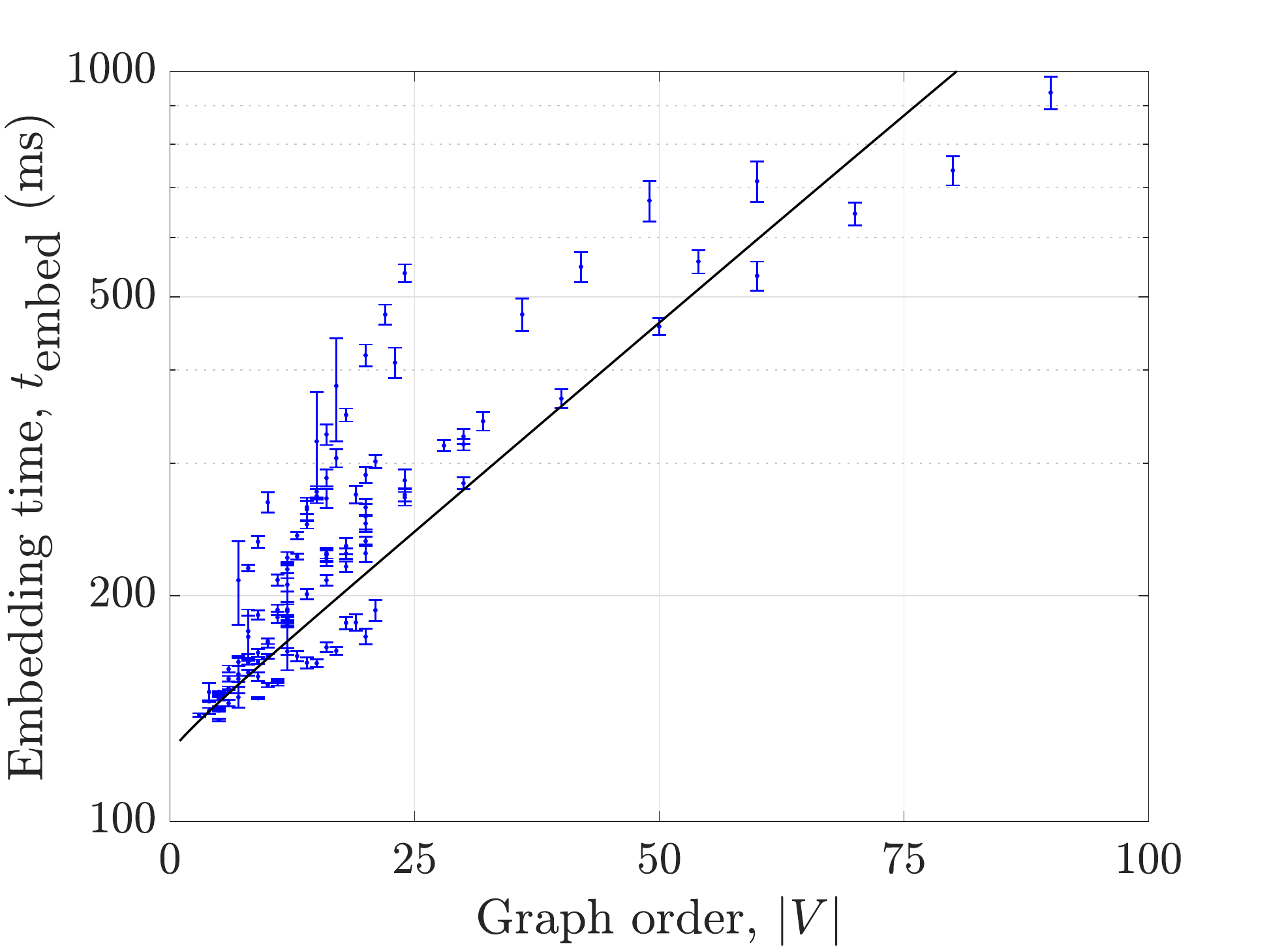}\\
\caption{Plot of graph order $|V|$ against the embedding time $t_\text{embed}$. Note the logarithmic scale in time.}\label{fig:embedding}
\end{center}
\end{figure}

First and foremost, from the results shown in Figure~\ref{fig:TH_v_Tstd_v_TC}(a) the extent of the advantage of the hybrid approach is  evident.
Indeed, this is to be expected given that, for a given DWMWIS problem, they differ (by definition) by $99\times t_\text{embed}$.
Although this might seem a trivial confirmation of this fact, the results help illustrate the extent of the advantage that the hybrid approach can have for such problems, a consequence of the absolute cost of the embedding.
This is visible in Figure~\ref{fig:embedding}, showing $t_\text{embed}$ as a function of the number of vertices in a graph.

From Figure~\ref{fig:TH_v_Tstd_v_TC}(b) it is also evident that no absolute quantum speedup was observed using the hybrid algorithm, and indeed there is a vast difference in scale between $T_C$ and $T_H$: the ``hardest'' problem was solved classically in less than 200ms, whereas the hybrid algorithm required almost 60 times as much time to solve it correctly.
The inability to observe any raw speedup is hardly surprising when one notes that, even if $k_{99}=1$ and $t_\text{embed}=t_\text{post}=0$, the fact that $t_\text{prog}\approx 20$ms means that that one would have $T_H > 2000$ms.
This programming time thus adds an essentially constant overhead, which would have less of an impact as larger problems (for which $k_{99}$ is much larger) become solvable.

\begin{figure}[h]
\begin{center}
\begin{tabular}{cc}
\includegraphics[width=0.5\textwidth]{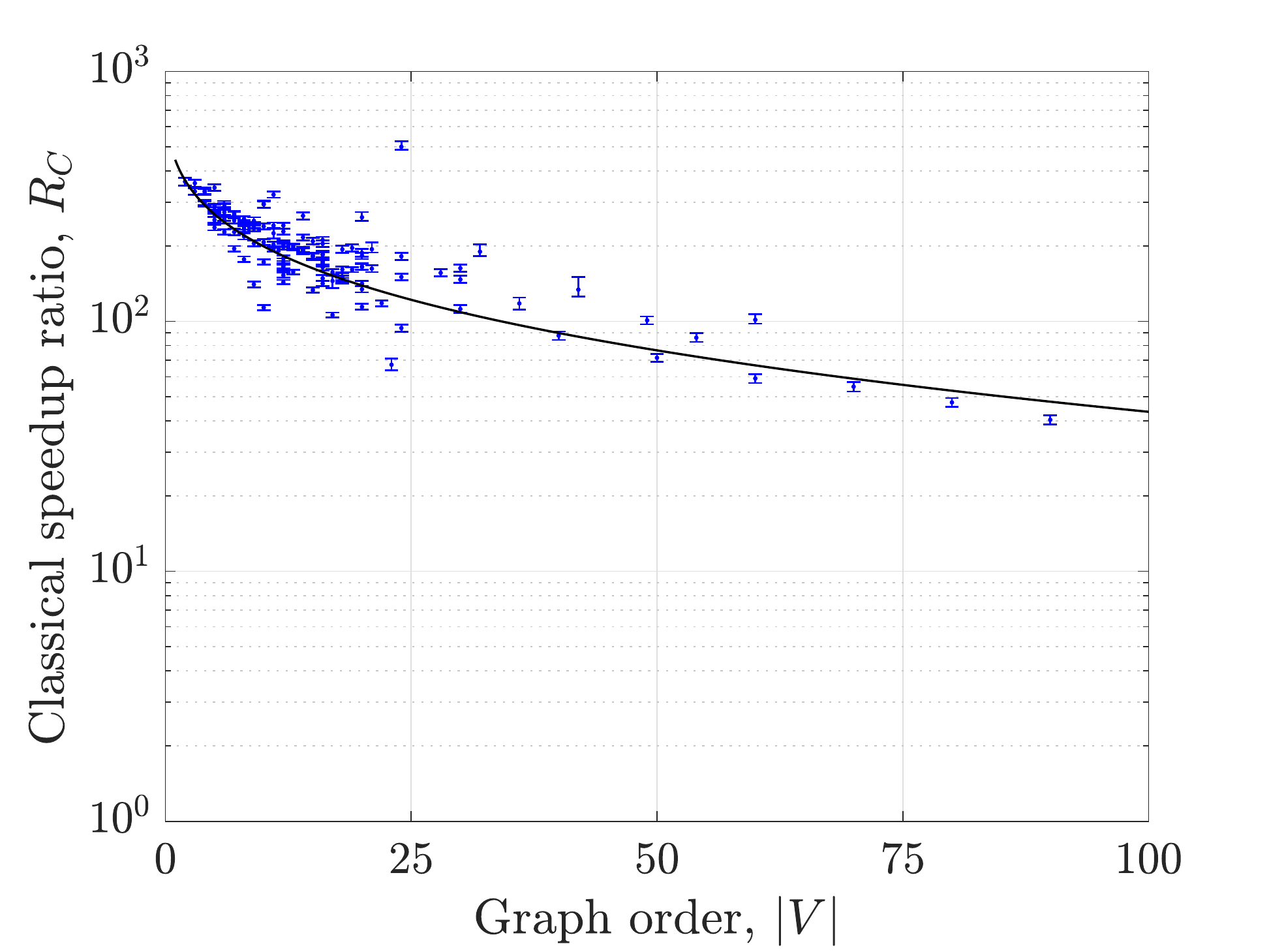} &\hspace{-5mm} \includegraphics[width=0.5\textwidth]{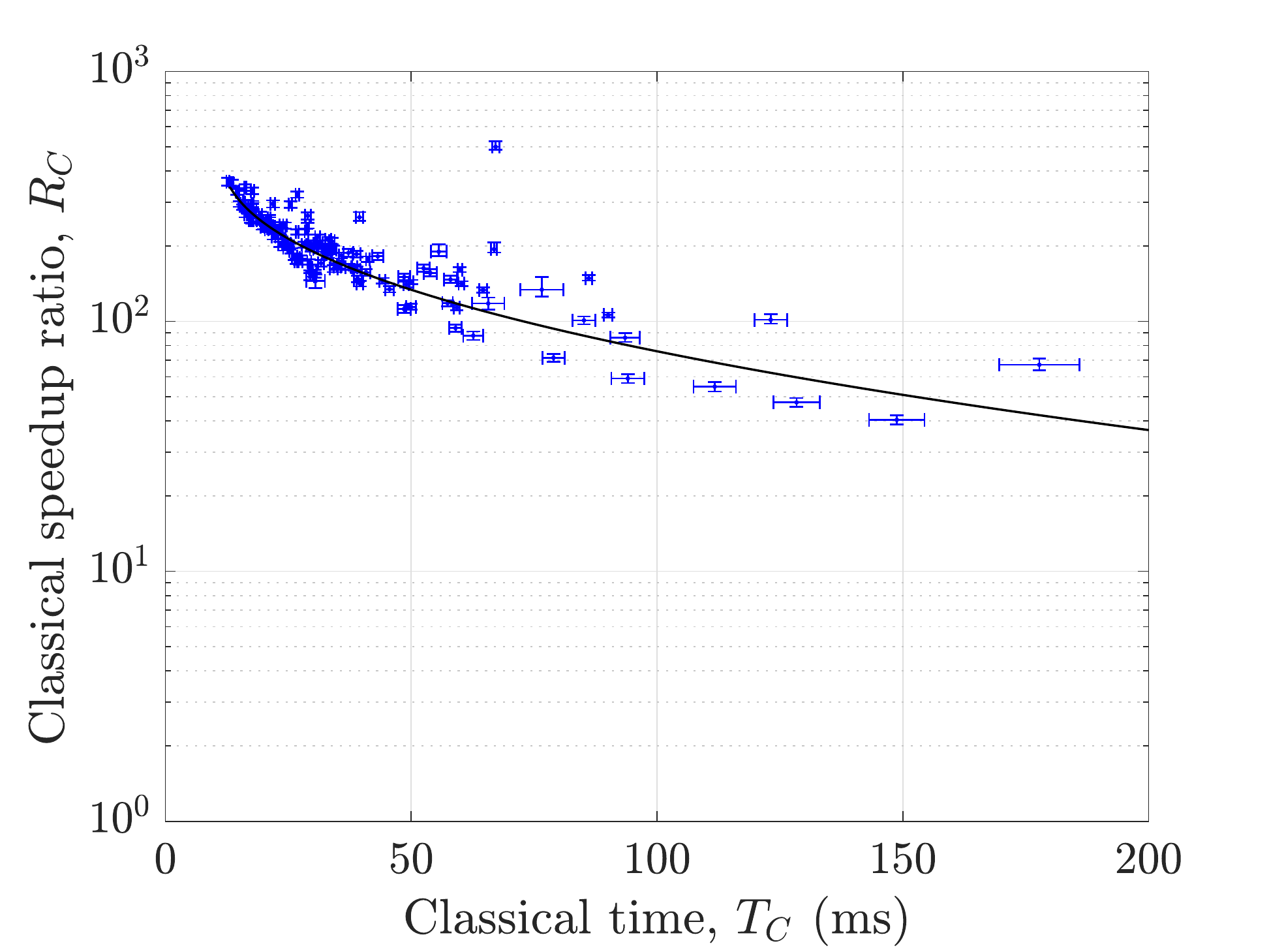}\\
(a) & (b)
\end{tabular}
\caption{Logarithmic plots of the scaling behaviour of the classical speedup ratio $R_C$ for the DWMWIS problem instances: (a) graph order $|V|$ against $R_C$; and (b) classical time $T_C$ against $R_C$.}
\label{fig:scalingOverall}
\end{center}
\end{figure}

\begin{figure}[h]
\begin{center}
\begin{tabular}{ccc}
\hspace{-3mm}\includegraphics[width=0.36\textwidth]{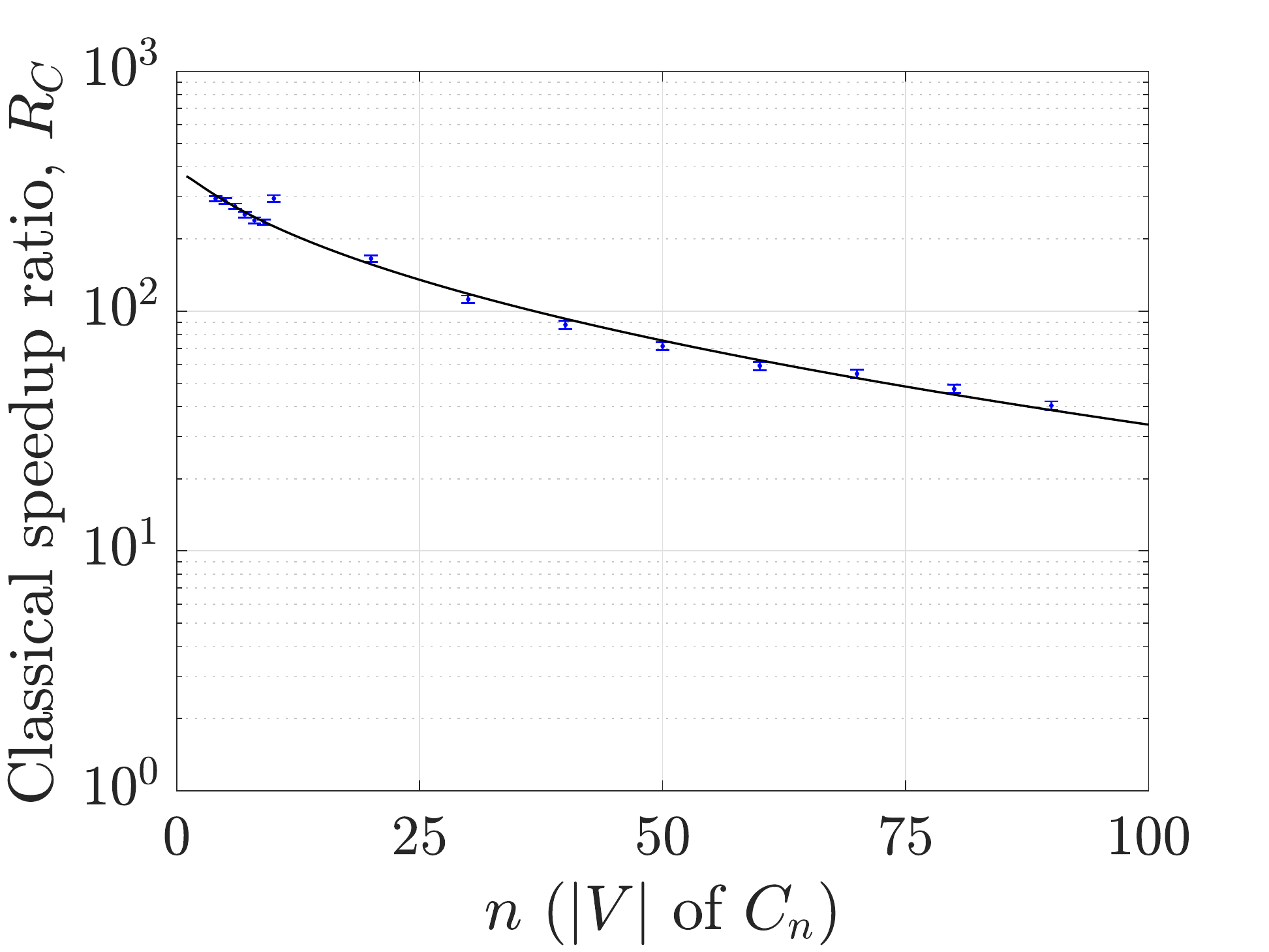} &\hspace{-10mm} \includegraphics[width=0.36\textwidth]{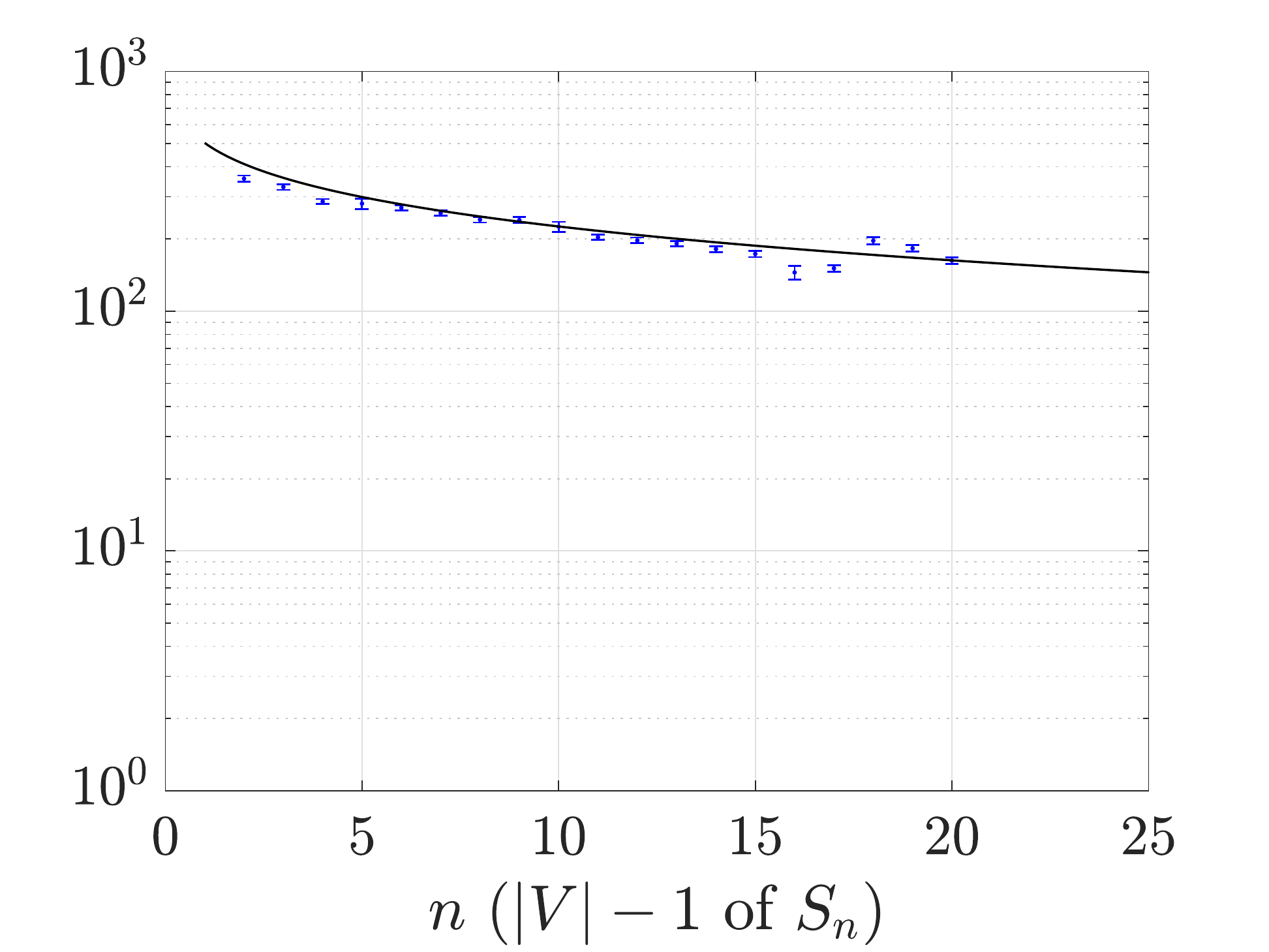} & \hspace{-10mm} \includegraphics[width=0.36\textwidth]{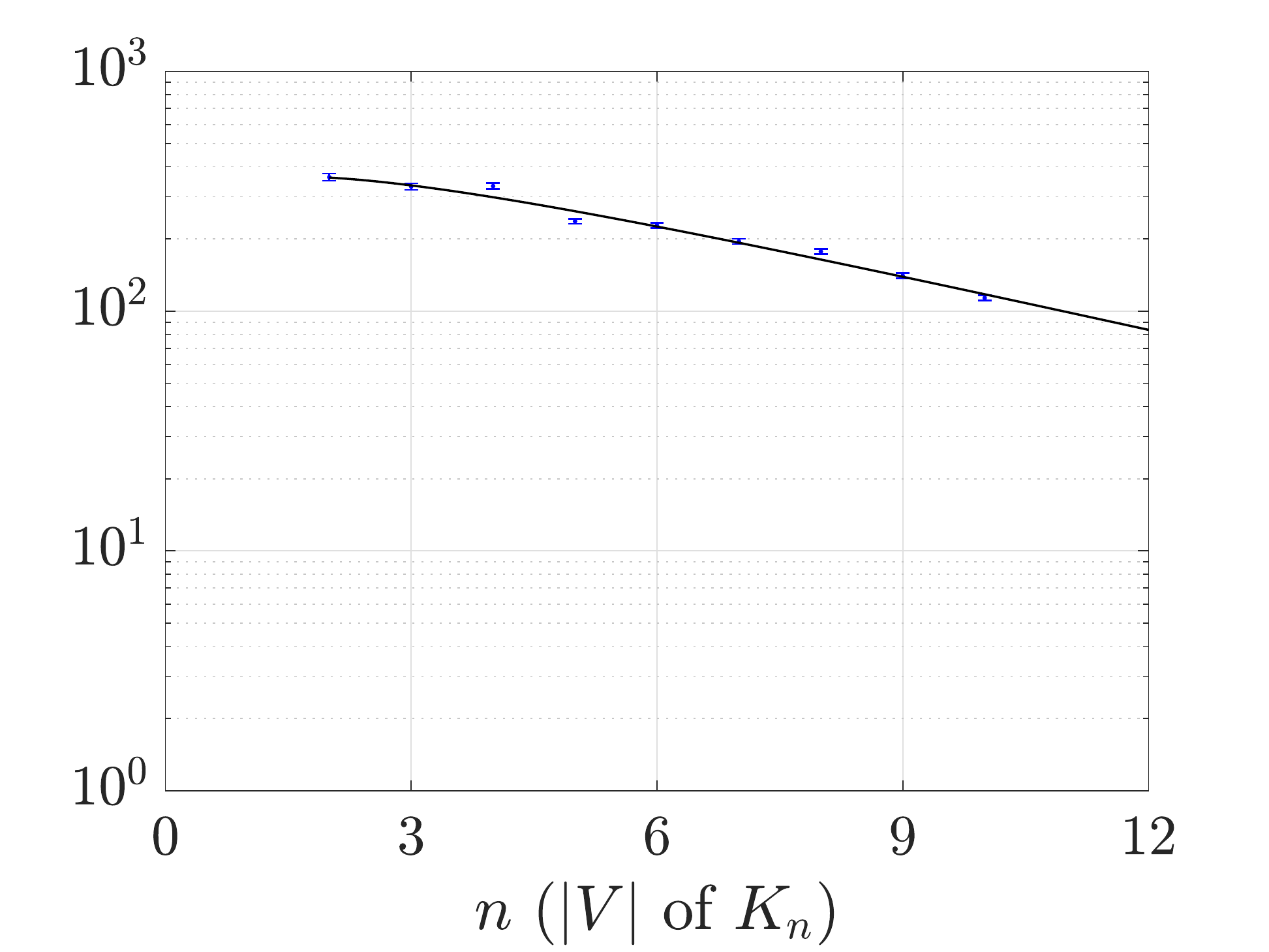}\\
(a) & (b) & (c)
\end{tabular}
\caption{Plots of the classical speedup ratio $R_C$ against $n$ for three families of graphs parameterised by $n$: (a) the $C_n$ graphs; (b) the $S_n$ graphs; (c) the $K_n$ graphs.}
\label{fig:scalingFamilies}
\end{center}
\end{figure}

Despite the absence of no overall speedup, it is interesting to examine the scaling behaviour of the hybrid approach, for which it will be useful to consider the ``classical speedup ratio'' $R_C=T_H/T_C$.
In Figure~\ref{fig:scalingOverall} we show the scaling behaviour of $R_C$ against two reasonable proxies of problem difficulty: the graph order $|V|$, which is proportional to the problem size, and the classical time $T_C$.
While there is much uncertainty in the exact nature of the scaling, this results indicate that the Hybrid algorithm has a better scaling behaviour than the classical algorithm.
This is more evident in Figure~\ref{fig:scalingFamilies}, where $R_C$ is plotted for specific graph families.
Thus, although no quantum speedup was found, the results leave open the possibility that such a speedup will be attainable in the future on larger devices with better control of the qubits, although many unknowns may plausibly alter the scaling behaviour in the future.

Nevertheless, the experiment was a successful proof-of-concept for the hybrid
paradigm we have presented.  In particular, the hybrid algorithm we implemented
provided large absolute gains over the standard quantum approach and showed
good scaling behaviour.  As larger and more efficient devices become available
and more problems of practical interest are studied, it will become clearer if/when
a quantum speedup might be obtainable in practise.

\section{Conclusion}

In this paper we presented a hybrid quantum-classical paradigm for quantum annealing algorithms aimed at countering the significant cost of the embedding process.
This approach is not only a hybrid paradigm but serves equally as a guide to identifying problems that may be amenable to quantum annealing.  
In particular, we identify those problems that require solving a large number of
related subproblems, each of which can be directed solved via annealing, may
permit a hybrid approach. This is obtained by reusing and modifying embeddings for the related
subproblems.

\clearpage

Our hybrid approach, along with its successful proof-of-principle implementation, sets 
the groundwork for addressing more complex problems of practical interest.
Choosing correctly suitable problems is a major step in finding practical uses for 
quantum computers in the near term future, and with deft choices, 
quantum speedups from hybrid approaches might soon be realisable.

\paragraph{Acknowledgements}
We thank  N.~Allen, C.~McGeoch, K.~Pudenz and S.~Reinhardt for fruitful discussions and critical comments.
This work has  been supported in part by the Quantum Computing Research Initiatives at Lockheed Martin.

\bibliographystyle{eptcs}
\bibliography{LHM}

\begin{thebibliography}{10}
\providecommand{\bibitemdeclare}[2]{}
\providecommand{\surnamestart}{}
\providecommand{\surnameend}{}
\providecommand{\urlprefix}{Available at }
\providecommand{\url}[1]{\texttt{#1}}
\providecommand{\href}[2]{\texttt{#2}}
\providecommand{\urlalt}[2]{\href{#1}{#2}}
\providecommand{\doi}[1]{doi:\urlalt{http://dx.doi.org/#1}{#1}}
\providecommand{\bibinfo}[2]{#2}

\bibitemdeclare{inproceedings}{Aaronson:2010aa}
\bibitem{Aaronson:2010aa}
\bibinfo{author}{S.~\surnamestart Aaronson\surnameend} (\bibinfo{year}{2010}):
  \emph{\bibinfo{title}{{BQP} and the polynomial hierarchy}}.
\newblock In: {\sl \bibinfo{booktitle}{STOC '10 Proceedings of the forty-second
  ACM symposium on Theory of computing}}, p. \bibinfo{pages}{141},
  \doi{10.1145/1806689.1806711}.

\bibitemdeclare{unpublished}{Abbott18}
\bibitem{Abbott18}
\bibinfo{author}{A.~A. \surnamestart Abbott\surnameend}, \bibinfo{author}{C.~S.
  \surnamestart Calude\surnameend}, \bibinfo{author}{M.~J. \surnamestart
  Dinneen\surnameend} \& \bibinfo{author}{R.~\surnamestart Hua\surnameend}
  (\bibinfo{year}{2018}): \emph{\bibinfo{title}{A Hybrid Quantum-Classical
  Paradigm to Mitigate Embedding Costs in Quantum Annealing}}.
\newblock
  \bibinfo{note}{\href{https://www.cs.auckland.ac.nz/research/groups/CDMTCS/researchreports/index.php?download&paper_file=684}{CDMTCS
  Research Report Series} 520}.

\bibitemdeclare{article}{Boixo:2014aa}
\bibitem{Boixo:2014aa}
\bibinfo{author}{S.~\surnamestart Boixo\surnameend}, \bibinfo{author}{T.~F.
  \surnamestart R{\o}nnow\surnameend}, \bibinfo{author}{S.~V. \surnamestart
  Isakov\surnameend}, \bibinfo{author}{Z.~\surnamestart Wang\surnameend},
  \bibinfo{author}{D.~\surnamestart Wecker\surnameend}, \bibinfo{author}{D.~A.
  \surnamestart Lidar\surnameend}, \bibinfo{author}{J.~M. \surnamestart
  Martinis\surnameend} \& \bibinfo{author}{M.~\surnamestart Troyer\surnameend}
  (\bibinfo{year}{2014}): \emph{\bibinfo{title}{Evidence for quantum annealing
  with more than one hundred qubits}}.
\newblock {\sl \bibinfo{journal}{Nat. Phys.}} \bibinfo{volume}{10}, p.
  \bibinfo{pages}{218}, \doi{10.1038/nphys2900}.

\bibitemdeclare{article}{Calude:2015aa}
\bibitem{Calude:2015aa}
\bibinfo{author}{C.~S. \surnamestart Calude\surnameend},
  \bibinfo{author}{E.~\surnamestart Calude\surnameend} \&
  \bibinfo{author}{M.~J. \surnamestart Dinneen\surnameend}
  (\bibinfo{year}{2015}): \emph{\bibinfo{title}{Adiabatic Quantum Computing
  Challenges}}.
\newblock {\sl \bibinfo{journal}{ACM SIGACT News}}
  \bibinfo{volume}{46}(\bibinfo{number}{1}), p.~\bibinfo{pages}{40},
  \doi{10.1145/2744447.2744459}.

\bibitemdeclare{incollection}{dwavebroadcast2016}
\bibitem{dwavebroadcast2016}
\bibinfo{author}{C.~S. \surnamestart Calude\surnameend} \&
  \bibinfo{author}{M.~J. \surnamestart Dinneen\surnameend}
  (\bibinfo{year}{2016}): \emph{\bibinfo{title}{Solving the Broadcast Time
  Problem Using a {D-Wave} Quantum Computer}}.
\newblock In \bibinfo{editor}{A.~\surnamestart Adamatzky\surnameend}, editor:
  {\sl \bibinfo{booktitle}{Advances in Unconventional Computing}},
  chapter~\bibinfo{chapter}{17}, {\sl \bibinfo{series}{Emergence, Complexity
  and Computation}}~\bibinfo{volume}{22}, \bibinfo{publisher}{Springer
  International}, \bibinfo{address}{Switzerland}, p. \bibinfo{pages}{439},
  \doi{10.1007/978-3-319-33924-5\_17}.

\bibitemdeclare{article}{Cho:2014aa}
\bibitem{Cho:2014aa}
\bibinfo{author}{A.~\surnamestart Cho\surnameend} (\bibinfo{year}{2014}):
  \emph{\bibinfo{title}{Quantum or not, controversial computer yields no
  speedup}}.
\newblock {\sl \bibinfo{journal}{Science}} \bibinfo{volume}{344}, p.
  \bibinfo{pages}{1330}, \doi{10.1126/science.344.6190.1330}.

\bibitemdeclare{article}{Choi:2008aa}
\bibitem{Choi:2008aa}
\bibinfo{author}{V.~\surnamestart Choi\surnameend} (\bibinfo{year}{2008}):
  \emph{\bibinfo{title}{Minor-embedding in adiabatic quantum computation: {I}.
  {T}he parameter setting problem}}.
\newblock {\sl \bibinfo{journal}{Quantum Inf. Processing}} \bibinfo{volume}{7},
  p. \bibinfo{pages}{193}, \doi{10.1007/s11128-008-0082-9}.

\bibitemdeclare{article}{DWave2X}
\bibitem{DWave2X}
\bibinfo{author}{\surnamestart {D-Wave Systems Inc.}\surnameend}
  (\bibinfo{year}{2016}): \emph{\bibinfo{title}{The {D-Wave 2X}\texttrademark{}
  Quantum Computer Technology Overview}}.
\newblock
  \urlprefix\url{http://www.dwavesys.com/sites/default/files/D-Wave%202X%20Tech%20Collateral_1016F_0.pdf}.

\bibitemdeclare{article}{DWave:postprocessing}
\bibitem{DWave:postprocessing}
\bibinfo{author}{\surnamestart {D-Wave Systems Inc.}\surnameend}
  (\bibinfo{year}{2016}): \emph{\bibinfo{title}{Postprocessing Methods on
  {D-Wave Systems}}}.
\newblock {\sl \bibinfo{journal}{Tech. Report Release 2.4 09-1105A-B}}.

\bibitemdeclare{article}{DWave:2013aa}
\bibitem{DWave:2013aa}
\bibinfo{author}{\surnamestart {D-Wave Systems Inc.}\surnameend}
  (\bibinfo{year}{2017}): \emph{\bibinfo{title}{Programming with {QUBO}s}}.
\newblock {\sl \bibinfo{journal}{Tech. Report Release 2.4 09-1002A-C}}.

\bibitemdeclare{article}{dwavesys2017}
\bibitem{dwavesys2017}
\bibinfo{author}{\surnamestart {D-Wave Systems Inc.}\surnameend}
  (\bibinfo{year}{2017}): \emph{\bibinfo{title}{{The D-Wave
  2000Q\texttrademark{} Quantum Computer Technology Overview}}}.
\newblock
  \urlprefix\url{https://www.dwavesys.com/sites/default/files/D-Wave%202000Q%20Tech%20Collateral_0117F2.pdf}.

\bibitemdeclare{article}{Denchev:2016aa}
\bibitem{Denchev:2016aa}
\bibinfo{author}{V.~S. \surnamestart Denchev\surnameend},
  \bibinfo{author}{S.~\surnamestart Boixo\surnameend}, \bibinfo{author}{S.~V.
  \surnamestart Isakov\surnameend}, \bibinfo{author}{N.~\surnamestart
  Ding\surnameend}, \bibinfo{author}{R.~\surnamestart Babbush\surnameend},
  \bibinfo{author}{V.~\surnamestart Smelyanskiy\surnameend},
  \bibinfo{author}{J.~\surnamestart Martinis\surnameend},
  \bibinfo{author}{\surnamestart \surnameend} \&
  \bibinfo{author}{H.~\surnamestart Neven\surnameend} (\bibinfo{year}{2016}):
  \emph{\bibinfo{title}{What is the Computational Value of Finite-Range
  Tunneling?}}
\newblock {\sl \bibinfo{journal}{Phys. Rev. X}} \bibinfo{volume}{6}, p.
  \bibinfo{pages}{031015}, \doi{10.1103/PhysRevX.6.031015}.

\bibitemdeclare{article}{Farhi:2000aa}
\bibitem{Farhi:2000aa}
\bibinfo{author}{E.~\surnamestart Farhi\surnameend},
  \bibinfo{author}{J.~\surnamestart Goldstone\surnameend},
  \bibinfo{author}{S.~\surnamestart Gutman\surnameend} \&
  \bibinfo{author}{M.~\surnamestart Sipser\surnameend} (\bibinfo{year}{2000}):
  \emph{\bibinfo{title}{Quantum Computation by Adiabatic Evolution}}.
\newblock {\sl
  \bibinfo{journal}{\href{https://arxiv.org/abs/quant-ph/0001106}{arXiv:quant-ph/0001106}}}.

\bibitemdeclare{book}{GJ79}
\bibitem{GJ79}
\bibinfo{author}{M.~R. \surnamestart Garey\surnameend} \&
  \bibinfo{author}{D.~S. \surnamestart Johnson\surnameend}
  (\bibinfo{year}{1979}): \emph{\bibinfo{title}{Computers and Intractability.
  {A} Guide to the Theory of {NP}-Completeness}}.
\newblock \bibinfo{publisher}{Freeman}, \bibinfo{address}{San Francisco}.

\bibitemdeclare{inproceedings}{Grover:1996aa}
\bibitem{Grover:1996aa}
\bibinfo{author}{L.~K. \surnamestart Grover\surnameend} (\bibinfo{year}{1996}):
  \emph{\bibinfo{title}{A fast quantum mechanical algorithm for database
  search}}.
\newblock In: {\sl \bibinfo{booktitle}{Proceedings, 28th Annual ACM Symposium
  on the Theory of Computing (STOC)}}, p. \bibinfo{pages}{212},
  \doi{10.1145/237814.237866}.

\bibitemdeclare{article}{Hen:2015aa}
\bibitem{Hen:2015aa}
\bibinfo{author}{I.~\surnamestart Hen\surnameend},
  \bibinfo{author}{J.~\surnamestart Job\surnameend},
  \bibinfo{author}{T.~\surnamestart Albash\surnameend}, \bibinfo{author}{T.~F.
  \surnamestart R{\o}nnow\surnameend}, \bibinfo{author}{M.~\surnamestart
  Troyer\surnameend} \& \bibinfo{author}{D.~A. \surnamestart Lidar\surnameend}
  (\bibinfo{year}{2015}): \emph{\bibinfo{title}{Probing for quantum speedup in
  spin-glass problems with planted solutions}}.
\newblock {\sl \bibinfo{journal}{Phys. Rev. A}} \bibinfo{volume}{92}, p.
  \bibinfo{pages}{042325}, \doi{10.1103/PhysRevA.92.042325}.

\bibitemdeclare{article}{Johnson:2011aa}
\bibitem{Johnson:2011aa}
\bibinfo{author}{M.~W. \surnamestart Johnson\surnameend},
  \bibinfo{author}{M.~H.~S. \surnamestart Amin\surnameend},
  \bibinfo{author}{S.~\surnamestart Gildert\surnameend},
  \bibinfo{author}{T.~\surnamestart Lanting\surnameend},
  \bibinfo{author}{F.~\surnamestart Hamze\surnameend},
  \bibinfo{author}{N.~\surnamestart Dickson\surnameend},
  \bibinfo{author}{R.~\surnamestart Harris\surnameend}, \bibinfo{author}{A.~J.
  \surnamestart Berkley\surnameend}, \bibinfo{author}{J.~\surnamestart
  Johansson\surnameend}, \bibinfo{author}{P.~\surnamestart Bunyk\surnameend},
  \bibinfo{author}{E.~M. \surnamestart Chapple\surnameend},
  \bibinfo{author}{C.~\surnamestart Enderud\surnameend}, \bibinfo{author}{J.~P.
  \surnamestart Hilton\surnameend}, \bibinfo{author}{K.~\surnamestart
  Karimi\surnameend}, \bibinfo{author}{E.~\surnamestart Ladizinsky\surnameend},
  \bibinfo{author}{N.~\surnamestart Ladizinsky\surnameend},
  \bibinfo{author}{T.~\surnamestart Oh\surnameend},
  \bibinfo{author}{I.~\surnamestart Perminov\surnameend},
  \bibinfo{author}{C.~\surnamestart Rich\surnameend}, \bibinfo{author}{M.~C.
  \surnamestart Thom\surnameend}, \bibinfo{author}{E.~\surnamestart
  Tolkacheva\surnameend}, \bibinfo{author}{C.~J.~S. \surnamestart
  Truncik\surnameend}, \bibinfo{author}{S.~\surnamestart Uchaikin\surnameend},
  \bibinfo{author}{J.~\surnamestart Wang\surnameend},
  \bibinfo{author}{B.~\surnamestart Wilson\surnameend} \&
  \bibinfo{author}{G.~\surnamestart Rose\surnameend} (\bibinfo{year}{2011}):
  \emph{\bibinfo{title}{Quantum annealing with manufactured spins}}.
\newblock {\sl \bibinfo{journal}{Nature}} \bibinfo{volume}{473}, p.
  \bibinfo{pages}{194}, \doi{10.1038/nature10012}.

\bibitemdeclare{article}{King:2015ab}
\bibitem{King:2015ab}
\bibinfo{author}{A.~D. \surnamestart King\surnameend},
  \bibinfo{author}{T.~\surnamestart Lanting\surnameend} \&
  \bibinfo{author}{R.~\surnamestart Harris\surnameend} (\bibinfo{year}{2015}):
  \emph{\bibinfo{title}{Performance of a quantum annealer on range-limited
  constraint satisfaction problems}}.
\newblock {\sl
  \bibinfo{journal}{\href{https://arxiv.org/abs/1502.02098}{arXiv:1502.02098
  [quant-ph]}}}.

\bibitemdeclare{article}{King:2014aa}
\bibitem{King:2014aa}
\bibinfo{author}{A.~D. \surnamestart King\surnameend} \& \bibinfo{author}{C.~C.
  \surnamestart McGeoch\surnameend} (\bibinfo{year}{2014}):
  \emph{\bibinfo{title}{Algorithm engineering for a quantum annealing
  platform}}.
\newblock {\sl
  \bibinfo{journal}{\href{https://arxiv.org/abs/1410.2628}{arXiv:1410.2628
  [cs.DS]}}}.

\bibitemdeclare{article}{King:2015aa}
\bibitem{King:2015aa}
\bibinfo{author}{J.~\surnamestart King\surnameend},
  \bibinfo{author}{S.~\surnamestart Yarkoni\surnameend}, \bibinfo{author}{M.~M.
  \surnamestart Nevisi\surnameend}, \bibinfo{author}{J.~P. \surnamestart
  Hilton\surnameend} \& \bibinfo{author}{C.~C. \surnamestart
  McGeoch\surnameend} (\bibinfo{year}{2015}):
  \emph{\bibinfo{title}{Benchmarking a quantum annealing processor with the
  time-to-target metric}}.
\newblock {\sl
  \bibinfo{journal}{\href{https://arxiv.org/abs/1508.05087}{arXiv:1508.05087
  [quant-ph]}}}.

\bibitemdeclare{article}{Ladd:2010aa}
\bibitem{Ladd:2010aa}
\bibinfo{author}{T.~D. \surnamestart Ladd\surnameend},
  \bibinfo{author}{F.~\surnamestart Jelezko\surnameend},
  \bibinfo{author}{R.~\surnamestart Laflamme\surnameend},
  \bibinfo{author}{Y.~\surnamestart Nakamura\surnameend},
  \bibinfo{author}{C.~\surnamestart Monroe\surnameend} \&
  \bibinfo{author}{J.~L. \surnamestart O'Brien\surnameend}
  (\bibinfo{year}{2010}): \emph{\bibinfo{title}{Quantum computers}}.
\newblock {\sl \bibinfo{journal}{Nature}} \bibinfo{volume}{464},
  p.~\bibinfo{pages}{45}, \doi{10.1038/nature08812}.

\bibitemdeclare{inproceedings}{Lanzagorta:2005aa}
\bibitem{Lanzagorta:2005aa}
\bibinfo{author}{M.~\surnamestart Lanzagorta\surnameend} \&
  \bibinfo{author}{J.~K. \surnamestart Uhlmann\surnameend}
  (\bibinfo{year}{2005}): \emph{\bibinfo{title}{Hybrid quantum-classical
  computing with applications to computer graphics}}.
\newblock In: {\sl \bibinfo{booktitle}{ACM SIGGRAPH 2005 Courses}},
  \bibinfo{series}{SIGGRAPH '05}, \bibinfo{publisher}{ACM},
  \bibinfo{address}{New York, NY}, \doi{10.1145/1198555.1198723}.

\bibitemdeclare{misc}{Marx10}
\bibitem{Marx10}
\bibinfo{author}{D.~\surnamestart Marx\surnameend} (\bibinfo{year}{2010}):
  \emph{\bibinfo{title}{Fixed parameter algorithms. {P}art 2: Treewidth}}.
\newblock \urlprefix\url{http://www.cs.bme.hu/~dmarx/papers/marx-warsaw-fpt2}.
\newblock \bibinfo{note}{Open lectures for PhD students in computer science,
  University of Warsaw, Poland}.

\bibitemdeclare{article}{McClean:2016aa}
\bibitem{McClean:2016aa}
\bibinfo{author}{J.~R. \surnamestart McClean\surnameend},
  \bibinfo{author}{J.~\surnamestart Romero\surnameend},
  \bibinfo{author}{R.~\surnamestart Babbush\surnameend} \&
  \bibinfo{author}{A.~\surnamestart Aspuru-Guzik\surnameend}
  (\bibinfo{year}{2016}): \emph{\bibinfo{title}{The theory of variational
  hybrid quantum-classical algorithms}}.
\newblock {\sl \bibinfo{journal}{New J. Phys.}} \bibinfo{volume}{18}, p.
  \bibinfo{pages}{023023}, \doi{10.1088/1367-2630/18/2/023023}.

\bibitemdeclare{inproceedings}{Pudenz:2016aa}
\bibitem{Pudenz:2016aa}
\bibinfo{author}{K.~L. \surnamestart Pudenz\surnameend} (\bibinfo{year}{2016}):
  \emph{\bibinfo{title}{Parameter Setting for Quantum Annealers}}.
\newblock In: {\sl \bibinfo{booktitle}{20th IEEE High Performance Embedded
  Computing Workshop Proceedings}}, \doi{10.1109/HPEC.2016.7761619}.

\bibitemdeclare{article}{Ronnow:2014aa}
\bibitem{Ronnow:2014aa}
\bibinfo{author}{T.~F. \surnamestart R{\o}nnow\surnameend},
  \bibinfo{author}{Z.~\surnamestart Wang\surnameend},
  \bibinfo{author}{J.~\surnamestart Job\surnameend},
  \bibinfo{author}{S.~\surnamestart Boixo\surnameend}, \bibinfo{author}{S.~V.
  \surnamestart Isakov\surnameend}, \bibinfo{author}{D.~\surnamestart
  Wecker\surnameend}, \bibinfo{author}{J.~M. \surnamestart
  Martinis\surnameend}, \bibinfo{author}{D.~A. \surnamestart Lidar\surnameend}
  \& \bibinfo{author}{M.~\surnamestart Troyer\surnameend}
  (\bibinfo{year}{2014}): \emph{\bibinfo{title}{Defining and detecting quantum
  speedup}}.
\newblock {\sl \bibinfo{journal}{Science}} \bibinfo{volume}{345}, p.
  \bibinfo{pages}{420}, \doi{10.1126/science.1252319}.

\bibitemdeclare{article}{Shin:2014aa}
\bibitem{Shin:2014aa}
\bibinfo{author}{S.~W. \surnamestart Shin\surnameend},
  \bibinfo{author}{G.~\surnamestart Smith\surnameend}, \bibinfo{author}{J.~A.
  \surnamestart Smolin\surnameend} \& \bibinfo{author}{U.~\surnamestart
  Vazirani\surnameend} (\bibinfo{year}{2014}): \emph{\bibinfo{title}{How
  ``Quantum'' is the {D-Wave} Machine?}}
\newblock {\sl
  \bibinfo{journal}{\href{https://arxiv.org/abs/1401.7087}{arXiv:1401.7087
  [quant-ph]}}}.

\bibitemdeclare{article}{sagemath}
\bibitem{sagemath}
\bibinfo{author}{\surnamestart {The Sage Developers}\surnameend}
  (\bibinfo{year}{2017}): \emph{\bibinfo{title}{{S}ageMath, the {S}age
  {M}athematics {S}oftware {S}ystem ({V}ersion 8.0)}}.
\newblock \urlprefix\url{http://www.sagemath.org}.

\bibitemdeclare{inproceedings}{Tran:2016aa}
\bibitem{Tran:2016aa}
\bibinfo{author}{T.~T. \surnamestart Tran\surnameend},
  \bibinfo{author}{M.~\surnamestart Do\surnameend}, \bibinfo{author}{E.~G.
  \surnamestart Rieffel\surnameend}, \bibinfo{author}{J.~\surnamestart
  Frank\surnameend}, \bibinfo{author}{Z.~\surnamestart Wang\surnameend},
  \bibinfo{author}{B.~\surnamestart O'Gorman\surnameend},
  \bibinfo{author}{D.~\surnamestart Venturelli\surnameend} \&
  \bibinfo{author}{J.~C. \surnamestart Beck\surnameend} (\bibinfo{year}{2016}):
  \emph{\bibinfo{title}{A Hybrid Quantum-Classical Approach to Solving
  Scheduling Problems}}.
\newblock In: {\sl \bibinfo{booktitle}{Proceedings of the Ninth International
  Symposium on Combinatorial Search}}, \bibinfo{publisher}{AAAI}.

\bibitemdeclare{article}{jobshopschedulingproblem2016}
\bibitem{jobshopschedulingproblem2016}
\bibinfo{author}{D.~\surnamestart Venturelli\surnameend},
  \bibinfo{author}{D.~J.~J. \surnamestart Marchand\surnameend} \&
  \bibinfo{author}{G.~\surnamestart Rojo\surnameend} (\bibinfo{year}{2015}):
  \emph{\bibinfo{title}{Job Shop Scheduling Solver based on Quantum
  Annealing}}.
\newblock {\sl
  \bibinfo{journal}{\href{https://arxiv.org/abs/1506.08479}{arXiv:1506.08479
  [quant-ph]}}}.

\bibitemdeclare{article}{Vinci:2015aa}
\bibitem{Vinci:2015aa}
\bibinfo{author}{W.~\surnamestart Vinci\surnameend},
  \bibinfo{author}{T.~\surnamestart Albash\surnameend},
  \bibinfo{author}{G.~\surnamestart Paz-Silva\surnameend},
  \bibinfo{author}{I.~\surnamestart Hen\surnameend} \& \bibinfo{author}{D.~A.
  \surnamestart Lidar\surnameend} (\bibinfo{year}{2015}):
  \emph{\bibinfo{title}{Quantum annealing correction with minor embedding}}.
\newblock {\sl \bibinfo{journal}{Phys. Rev. A}} \bibinfo{volume}{92}, p.
  \bibinfo{pages}{042310}, \doi{10.1103/PhysRevA.92.042310}.

\bibitemdeclare{article}{maxindepsetqubo2017}
\bibitem{maxindepsetqubo2017}
\bibinfo{author}{S.~\surnamestart Yarkoni\surnameend},
  \bibinfo{author}{A.~\surnamestart Plaat\surnameend} \&
  \bibinfo{author}{T.~\surnamestart B\"{a}ck\surnameend}
  (\bibinfo{year}{2017}): \emph{\bibinfo{title}{First results solving
  arbitrarily structured Maximum Independent Set problems using quantum
  annealing}}.
\newblock
  \urlprefix\url{http://liacs.leidenuniv.nl/~plaata1/papers/MIS_yarkoni.pdf}.

\end{thebibliography}

\end{document}